\documentclass[a4paper]{scrartcl}
\usepackage{ifthen}
\usepackage{epsfig}
\usepackage{amssymb}
\usepackage{graphicx}
\usepackage{amsmath}
\usepackage{array}
\usepackage[center]{subfigure}

\newcommand{\ba}{\begin{eqnarray}}
\newcommand{\ea}{\end{eqnarray}}
\newcommand{\be}{\begin{equation}}
\newcommand{\ee}{\end{equation}}
\newcommand{\fplq}{$f_{B\pi}^+(q^2)~$}
\newcommand{\fzeroq}{$f_{B\pi}^0(q^2)~$}
\newcommand{\fpl}{$f_{B\pi}^+~$}
\newcommand{\fzero}{$f_{B\pi}^0~$}
\newcommand{\B}{\bar{B}^0}
\newcommand{\GeV}{~\mbox{GeV}}

\begin{document}
\begin{titlepage}
\vfill
\begin{center}
{\Large\bf
$B\!\to\! \pi \ell \nu_\ell$ Width and $|V_{ub}|$ \\[2mm] from
QCD Light-Cone Sum Rules }\\[2cm]
{\large\bf
A.~Khodjamirian\,$^{(a)}$,
Th.~Mannel\,$^{(a)}$, N.~Offen\,$^{(b)}$ and Y.-M. Wang $^{(a)}$}\\[0.5cm]
{\it  $^{(a)}$\,Theoretische Physik 1, Physik Department,
Universit\"at Siegen,\\ D-57068 Siegen, Germany }\\[2mm]
{\it $^{(b)}$ Institut f\"ur Theoretische Physik,  Universit\"at Regensburg, \\
D-93040 Regensburg, Germany  }
\end{center}
\vfill
\begin{abstract}
We employ the $B\to\pi$  form factors obtained from  QCD
light-cone sum rules and calculate the $B\to \pi \ell \nu_\ell$
width ($\ell=e,\mu$) in units of $1/|V_{ub}|^2$, integrated over
the region  of accessible momentum transfers, $0\leq q^2\leq 12.0
~\mbox{GeV}^2$. Using the most recent BABAR-collaboration
measurements we extract
$|V_{ub}|=(3.50^{+0.38}_{-0.33}\big|_{th.}\pm 0.11
\big|_{exp.})\times 10^{-3}$. The sum rule results for the form
factors, taken as an input for a $z$-series parameterization,
yield the $q^2$-shape in the whole semileptonic region of $B\to
\pi\ell\nu_\ell$. We also present the charged lepton energy
spectrum in this decay. Furthermore, the current situation with
$B\to \tau\nu_\tau$ is discussed from the QCD point of view. We
suggest to use the ratio of the $B\to \pi \tau\nu_\tau$ and $B\to
\pi\ell \nu_\ell ~(\ell =\mu,e) $ widths as an additional test of
Standard Model. The sensitivity of this observable to new physics
is illustrated by including a charged Higgs-boson contribution in
the semileptonic decay amplitude.\\


\end{abstract}
\vfill
\end{titlepage}

\section{Introduction}
Currently, there is a tension
between the two values of $|V_{ub}|$ extracted
from  inclusive and exclusive semileptonic $B$-decays
involving $b\to u$ transition.
While the inclusive analyses
typically yield a central value of $|V_{ub}|$ larger than
$4  \times 10^{-3}$, the exclusive determinations produce central values well
below this. This tension is not significant; it ranges at a level of
3$\sigma$, but it already has created a significant amount of speculations
concerning possible new physics effects.
This is in contrast to the situation with
$|V_{cb}|$, where both the inclusive  as well as the exclusive
determinations yield consistent values with an
uncertainty of roughly 2\% (for a review on $|V_{cb}|$ and $|V_{ub}|$ see \cite{vubvcb}).

The theoretical description of inclusive semileptonic $B$ decays relies on the
heavy-quark expansion which has reached a mature state.
Still, the situation with $b\to u$ inclusive decays is more complicated than with the
dominant $b\to c$ ones.
In order to suppress the charm background in the
inclusive $b\to u$ decays, severe phase space cuts are necessary,
for which most theoretical methods
cannot rely on  the heavy quark expansion based on the local operator-product expansion.
The heavy quark expansion for this case uses non-local matrix elements corresponding to the light-cone distribution functions of the $B$ meson,
which also appear e.g., in the radiative $b \to s$ decays.
Due to this more complicated structure of the expansion, it is quite hard
to estimate subleading terms in the heavy-quark expansion for $b\to u$
decays. In particular the inclusive method  has been scrutinized for a
missing systematic effect like e.g., ``weak annihilation'', however, no missing
pieces could be identified yet.
Still, it is believed that this method allows a determination of $|V_{ub}|$
at a precision of a little better than 10\%.

As far as exclusive decays are concerned, heavy-quark symmetries
restrict the form factors for the heavy-light $b \to u$
transitions much weaker  than the ones for the $b \to c$
transitions. Hence, the determination of $|V_{ub}|$ from exclusive
decays such as $B \to \pi \ell \bar{\nu}$ and $B \to \rho \ell
\bar{\nu}$ requires a QCD calculation of the relevant hadronic
form factors. State-of-the-art calculations do not rely on quark
models any more, since the latter cannot be directly related to
QCD. Already for many years, the form factors are obtained, on one
hand, from lattice simulations and, on the other hand, from QCD
sum rules. The two approaches are complementary: While the lattice
techniques can calculate close to the maximal leptonic momentum
transfer $q^2$, the QCD sum rule approach works best for small
$q^2$. Extrapolating both predictions to the full phase space
yields consistent results and hence there is some confidence that
the form factors, in particular, for $B \to \pi \ell \bar{\nu}$
are known with an uncertainty of 10-15\%.

Currently, $B\to \pi l\nu_\ell$ is the most reliable exclusive
channel to extract $|V_{ub}|$. There is a steady progress in
measuring the branching fraction and $q^2$-distribution for
$l=\mu,e$ (see \cite{BaBarnew1,Babarnew2,Belle} for the latest
results). The hadronic vector form factor $f^+_{B\pi}(q^2)$ and
its scalar counterpart $f^0_{B\pi}(q^2)$ relevant for this decay
are defined as \footnote{Throughout this paper, we assume isospin
symmetry for $B^0$ and $B^\pm$ semileptonic decays  and consider
the $\bar{B}^0\to \pi^+\ell^-\bar{\nu}_\ell$ mode for
definiteness, denoting it as $B\to \pi \ell\nu_\ell$ for brevity.}
\ba \langle\pi^+(p)|\bar{u} \gamma_\mu b |\B(p+q)\rangle&=&
f^+_{B\pi}(q^2)\Big[2p_\mu +
\left(1-\frac{m_B^2-m_\pi^2}{q^2}\right) q_\mu\Big] \nonumber
\\
&+&f^0_{B\pi}(q^2)\frac{m_B^2-m_\pi^2}{q^2}q_\mu,
\label{eq:fplBpi} \ea where $f^+_{B\pi}(0)=f^0_{B\pi}(0)$. The most recent lattice QCD computations
with three dynamical flavours
\cite{HPQCD,FermilabMILC} predict these form factors
at $q^2\geq  16$ GeV$^2$, in the upper part of the
semileptonic region $0\leq q^2 \leq (m_B-m_\pi)^2\simeq 26.4 \GeV^2$, with an accuracy reaching $10\%$.
There are also recent results available \cite{QCDSF} in the quenched  approximation on a fine lattice.
QCD light-cone sum rules (LCSR) with pion distribution amplitudes  (DA's)
allow one to calculate the $B\to \pi$ form factors
\cite{Khodjamirian:1997ub,Bagan:1997bp,Ball:1998tj,Ball:2004ye,DKMMO}
at small and intermediate momentum transfers, $0\leq q^2 \leq q^2_{max}$, where
the choice of $q^2_{max}$ varies between  12 and 16 GeV$^2$.

The main goal of this paper is to present an updated LCSR
prediction for the width of $B\to \pi\ell \nu_\ell$ ($\ell=e,\mu$)
which is then used to extract $|V_{ub}|$. This work complements
\cite{DKMMO} where LCSR for \fpl and \fzero were rederived,
employing the $\overline{MS}$ scheme for the virtual $b$-quark in
the correlation function. In \cite{DKMMO}, the shape of the form
factor \fplq predicted from LCSR was fitted to the earlier BABAR
measurement \cite{BABAR07} of the $q^2$-distribution in $B\to \pi
l \nu_\ell$. In this way, some input parameters of LCSR were
constrained, allowing one to decrease the theoretical uncertainty
of the value $f^+_{B\pi}(0)$, the main prediction of \cite{DKMMO}.

In this paper, we follow a different strategy. The intervals of
the Gegenbauer moments of the pion twist-2 DA are constrained
using the LCSR for the pion electromagnetic (e.m.) form factor at
spacelike momentum transfers. The calculated form factor is then
fitted to the available experimental data on this form factor. We
also slightly update the other input parameters, and recalculate
the form factors \fplq and \fzeroq at  $0\leq q^2<q^2_{max}$  from
LCSR. Our main prediction is the integral: \be \Delta\zeta\,(0,
q_{max}^2)\equiv
\frac{G_F^2}{24\pi^3}\int\limits_0^{q_{max}^2}dq^2p_\pi^3
|f_{B\pi}^+(q^2)|^2= \frac{1}{|V_{ub}|^2\tau_{B^0}}
\int\limits_0^{q_{max}^2} dq^2\frac{d{\cal B}(B\to \pi\ell
\nu_\ell)}{dq^2}\,, \label{eq:zeta} \ee where
$p_\pi=\sqrt{(m_B^2+m_\pi^2-q^2)^2/4m_B^2-m_\pi^2}$ is the pion
3-momentum in the $B$-meson rest frame, and the above equation is
valid for $\ell=e,\mu$ in the limit $m_l=0$. As in \cite{DKMMO},
the value $q^2_{max}=12.0$ GeV$^2$ is adopted. The predicted
$\Delta\zeta (0,12\GeV^2)$  is used to extract $|V_{ub}|$ from the
most recent BABAR-collaboration results \cite{BaBarnew1,Babarnew2}
for the measured partial branching fraction integrated over the
same $q^2$-region. Furthermore, we predict the form factors in the
whole semileptonic region by fitting the  LCSR results at $q^2\leq
q^2_{max}$ to the $z$-series parameterization in the form
suggested in \cite{BCL}. In addition to the $q^2$-distribution of
the width, we present a ``by-product'' observable: the lepton
energy spectrum in $B \to \pi\ell \nu_\ell$. Finally, we calculate
the ratio of the $B\to \pi \tau\nu_\tau$ and $B\to \pi\ell
\nu_\ell$ ($\ell =e,\mu)$ widths, which is independent of
$|V_{ub}|$ and in Standard Model (SM) is fully determined by the
ratio \fzeroq/\fplq\!. We also discuss the current tension between
the SM prediction and measurements of the leptonic
$B\to\tau\nu_\tau$ width, and suggest to use the semileptonic
decay $B\to \pi \tau\nu_\tau$ decay, originating  from the same
flavour-changing interaction, as an additional indicator of new
physics. As an illustrative example, we consider the influence of
a charged Higgs-boson exchange on the $B\to \pi \tau\nu_\tau$
width.

In what follows, a brief outline
of the LCSR method for $B\to \pi$ form factors is given in Sect.~2.
The choice of the
Gegenbauer moments of the pion DA is discussed in Sect.~3.
In Sect.~4 we present
our numerical results for the form factors and for the
integrated width. In Sect. 5 we discuss the $z$-series parameterization
and present our predictions for the whole semileptonic region.
In Sect. 6 we discuss the current situation with $B\to \tau\nu_\tau$
and the decay $B\to \pi \tau\nu_\tau$,
concluding in Sect.~7.

\section{ Outline of the method and input}
To obtain the LCSR for the form factors \fpl and \fzero
one uses the correlation function
\ba
F_{\mu}(p,q)=i\int d^4x ~e^{i q\cdot x}
\langle \pi^+(p)|T\left\{\bar{u}(x)\gamma_\mu b(x),
m_b\bar{b}(0)i\gamma_5 d(0)
\right\}|0\rangle
\nonumber\\
=F(q^2,(p+q)^2)p_\mu +\widetilde{F}(q^2,(p+q)^2)q_\mu\,,
\label{eq:corr}
\ea
of the $b\to u$ vector current and the $B$-meson interpolating current.
As explained e.g., in \cite{DKMMO}, the product
of the quark operators in the above
is expanded near the light-cone, provided both external momenta
are highly-virtual: $(p+q)^2, q^2\ll m_b^2$. The
operator-product expansion (OPE)
result for the  invariant amplitudes in (\ref{eq:corr})
is obtained in a (schematic) form:
\ba
F^{OPE}(q^2,(p+q)^2)=\sum\limits_{t}\int {\cal D}u_i~ T^{(t)}(q^2,(p+q)^2,u_i,\bar{m}_b,\alpha_s,\mu_f)
\varphi_\pi^{(t)}(u_i,\mu_f)\,,
\label{eq:OPE}
\ea
with  a similar expression for $\widetilde{F}^{(OPE)}$. In the above,
the pion light-cone DA's $\varphi^{(t)}(u_i)$ of growing
twist $t=2,3,4$ are defined as functions of  the light-cone
momentum fractions $u_i$; for the two-particle DA's $u_1=u$,
$u_2=1-u\equiv
\bar{u}$, with the integration over $u$. The DA's are convoluted with
the coefficient functions
(hard-scattering amplitudes) $T^{(t)}$ and $\widetilde{T}^{(t)}$
at the factorization scale $\mu_f$. The currently accessible
approximation for the light-cone OPE includes the contributions
of all two- and three-particle DA's up to the twist 4.
For the leading twist-2 and twist-3  contributions the coefficient functions
are calculated in NLO, taking into account the  $O(\alpha_s)$  gluon radiative corrections. The input in the OPE (\ref{eq:OPE}) includes: the $b$-quark mass
(in the $\overline{MS}$ scheme)  and QCD coupling in the coefficient  functions,
as well as the parameters of the universal pion DA's.

The dispersion relation for the correlation function (\ref{eq:corr}) in the channel of the $B$-meson interpolating current with momentum $p+q$ is then employed to access the form factors:
\be
F^{OPE}(q^2,(p+q)^2)=\frac{2f_Bm_B^2f^+_{B\pi}(q^2)}{m_B^2-(p+q)^2}+...,
\label{eq:dispF}
\ee
\ba
\widetilde{F}^{OPE}(q^2,(p+q)^2)=\frac{f_Bm_B^2}{m_B^2-(p+q)^2}\Big[
f^+_{B\pi}(q^2)\left(1-\frac{m_B^2-m_\pi^2}{q^2}\right)
\nonumber \\
+ f^0_{B\pi}(q^2)\frac{m_B^2-m_\pi^2}{q^2} \Big]+...,
\label{eq:disptF} \ea
where $f_B=\langle
\bar{B}|m_b\bar{b}i\gamma_5 d|0\rangle/m_B^2$ is the $B$-meson
decay constant. The ellipses in the above relations indicate the
integrals over the spectral densities of excited and continuum
$B$-states, for which the quark-hadron duality ansatz is used.
More specifically, one approximates the higher-state contributions
in (\ref{eq:dispF}) and (\ref{eq:disptF}) by the integrals over
the spectral density of the calculated invariant amplitudes
$Im_{(p+q)^2}F^{OPE}(q^2,s)$ and
$Im_{(p+q)^2}\widetilde{F}^{OPE}(q^2,s)$, respectively. This
approximation brings the effective threshold $s_0^B$ into
play. The final form of LCSR is obtained after applying the Borel
transformation to (\ref{eq:dispF}) and (\ref{eq:disptF}) replacing
the variable $(p+q)^2$ by the Borel parameter $M^2$; e.g., for the form
factor \fpl one obtains: \be
f^+_{B\pi}(q^2)=\left(\frac{e^{m_B^2/M^2}}{2m_B^2f_B}\right)\frac1{\pi}\int\limits_{m_b^2}^{s_0^B}ds~Im_{(p+q)^2}F^{OPE}(q^2,s)e^{-s/M^2}\,.
\label{eq:LCSR} \ee The second LCSR obtained from
(\ref{eq:disptF}) and combined with (\ref{eq:LCSR}) allows one to
calculate the form factor \fzeroq\!. Both sum rules are reliable
up to $q^2_{max}\sim m_B^2-2m_b\chi$, where $\chi$ is some large
scale, independent of $m_b$, so that at $q^2\leq q^2_{max}$ the
truncated light-cone  OPE can be trusted.

The interval  of $M^2$  in (\ref{eq:LCSR}) is constrained by
combining the two usual criteria for a QCD sum rule: smallness of the
power corrections (here the contributions of three-particle and twist-4
DA's to $F^{OPE}$), and, simultaneously, a moderate magnitude
of the hadronic continuum contribution.
The interval of $s_0^B$ is constrained by equating
the $B$-meson mass calculated from LCSR to its
experimental value. Finally, the decay constant $f_B$ is
calculated from the two-point sum rule with the same $\alpha_s$ accuracy.
The explicit expressions for the amplitudes $F^{OPE}$, $\widetilde{F}^{OPE}$
and their spectral functions entering LCSR, as well as a
detailed description of all pion DA's entering
(\ref{eq:OPE}), can be found in \cite{DKMMO}.

The most important contributions to the LCSR (\ref{eq:LCSR})
originate from the twist-2 and twist-3 terms in the OPE  (\ref{eq:OPE}).
The twist-2 pion DA $\varphi^{(2)}_\pi(u_1,u_2,\mu)=f_\pi\varphi_\pi(u,\mu)$,
is normalized to the pion decay constant $f_\pi$. The shape
of $\varphi_\pi(u,\mu)$ is determined by
the coefficients of the Gegenbauer-polynomial expansion (Gegenbauer moments),
to  be discussed in the next section. In the twist-3 pion DA's,
the most important input parameter is
the normalization coefficient $\mu_\pi=m_\pi^2/(m_u+m_d)$
related to the quark-condensate density.
The parameters determining the shapes of the twist-3,4 DA's
are known with a sufficient accuracy from the two-point QCD sum rules
(see e.g., \cite{BBL}).

\section{ Gegenbauer moments from the pion e.m. form factor}

For the twist-2 pion DA
we use the same approximation  as in \cite{DKMMO},
\be
\varphi_\pi(u,\mu_f)=6u\bar{u}\Big(1 +a_2^\pi(\mu_f)C_2^{3/2}(u-\bar{u})
+a_4^\pi(\mu_f)C_4^{3/2}(u-\bar{u})\Big)\,,
\label{eq:tw2DA}
\ee
retaining  the two
nonvanishing Gegenbauer moments $a_2^\pi$ and $a_4^\pi$ and neglecting all
higher moments, so that $a_{>4}^\pi=0$. This
approximation is justified because the renormalization
suppresses higher Gegenbauer moments at relatively large scales
$\mu_f$, typical for the LCSR (\ref{eq:LCSR}).
The uncertainties of the input values of $a_{2,4}^\pi(1 \mbox{GeV})$ are larger
than  very small effects of  the NLO evolution, which we also neglect.

In \cite{DKMMO} these two parameters
were constrained  by fitting the $B\to \pi$ form factor
calculated from LCSR at different
$q^2$ to the measured shape of $B\to \pi \ell \nu_{\ell}$,
yielding $a_2^\pi(1 \mbox{GeV})=0.16\pm 0.01$ and
~$a_4^\pi(1 \mbox{GeV})=0.04\pm 0.01$, where the uncertainties
only take into account the experimental error in the shape.

Here  we refrain from using the $B\to\pi \ell \nu_\ell$ data and
obtain an independent  constraint on the two  Gegenbauer moments
employing the pion e.m. form factor $F_\pi(Q^2)$ in the spacelike
region, defined as \be
 \langle \pi(p+q)|j_\mu^{em}|\pi(p)\rangle=
(2p+q)_\mu F_\pi(Q^2)\,,
\label{eq:fpi}
\ee
where  $j_\mu^{em}=\frac23\bar{u}(x)\gamma_\mu u (x) - \frac13\bar{d}(x)\gamma_\mu d (x)$ and  $Q^2=-q^2$.
The LCSR for $F_\pi(Q^2)$ derived in
\cite{BH,BKM} and updated in \cite{BijnAK}
is based on the correlation function, similar to (\ref{eq:corr}),
with virtual $u,d$ quarks instead of the $b$-quark and with
the axial-vector current instead of the pseudoscalar current.
This sum rule has NLO accuracy in the leading twist-2 term, with
the nonleading terms up to twist-6 taken into account. The $O(\alpha_s/Q^2)$ term
in LCSR correctly reproduces
the large-$Q^2$  QCD asymptotics of the pion form factor, whereas the soft
contributions dominate in the intermediate $Q^2$ region.
Importantly, the twist-3 contribution to the LCSR for $F_\pi(Q^2)$
vanishes in the chiral limit. Altogether, the sum rule for the pion e.m. form factor
is more sensitive  to the twist-2 pion DA
than the sum rules for heavy-to-light form factors.
The LCSR for  $F_\pi(Q^2)$ with the currently
achieved accuracy is applicable
at intermediate $Q^2$, from $O(1\mbox{GeV}^2)$ to a few GeV$^2$;
in the same region this form factor was accurately measured by the JLab experiment \cite{JLAB}. Preliminary comparison of the LCSR with
these data was done in \cite{talk09}. With the pion DA
given by (\ref{eq:tw2DA}) and taking the remaining
input from \,\cite{BijnAK}\, we recalculated
the pion e.m. form factor at $Q^2$ in the region up to a few GeV$^2$
as a function
of $a_2^\pi(1\mbox{GeV}^2)$ and $a_4^\pi(1\mbox{GeV}^2)$.
As shown in Fig.~\ref{fig:ffpion}, the result was fitted to
the seven data points  at 0.6 GeV$^2\leq Q^2\leq 2.45$~GeV$^2$,
presented in \cite{JLAB}, yielding:
\begin{figure}[t!]
\centering
\includegraphics[scale=0.8]{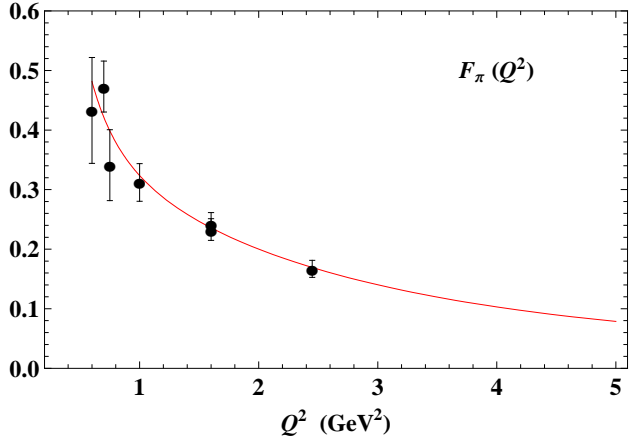}
\vspace{-0.3cm} \caption{\it The pion e.m.
form factor calculated
from LCSR \cite{BKM,BijnAK}
as a  function of Gegenbauer moments $a^\pi_2(1\GeV)$ and $a^\pi_4(1\GeV)$
and fitted (solid) to the experimental data points taken from
from \cite{JLAB}.}
\label{fig:ffpion}
\end{figure}
\be a_2^\pi(\mbox{1 GeV })=0.17\pm 0.08,
~~ a_4^\pi(\mbox{1 GeV })=0.06\pm 0.10\,,
\label{eq:a2a4}
\ee
where the uncertainties include experimental errors
and the variation of other input parameters taken as in \cite{BijnAK}.
We adopt these intervals for the
numerical analysis of the sum rule
(\ref{eq:LCSR}), neglecting the correlation of the two
uncertainties in (\ref{eq:a2a4}).
Note that the value of $a_2^\pi$ presented above is consistent with the
direct calculations of this parameter in lattice QCD \cite{lata2} and from
QCD sum rules \cite{BBL,KMM}.

The  $\gamma^*\gamma\to \pi^0$ transition form factor,
where the virtual photon has a spacelike virtuality $Q^2$  is
another important hadronic matrix element
that depends on the properties of the pion DA $\varphi_\pi(u)$.
The method \cite{pigamma} of combining LCSR with  the dispersion relation
in the photon virtuality predicts this form factor
starting from $Q^2\sim 1\,\mbox{GeV}^2$. It is important that,
according to the sum rule approach, the photon-pion transition form factor
contains a considerable nonperturbative soft contribution.
The calculation \cite{pigamma} was reconsidered in \cite{talk09}
with the same intervals of $a_2^\pi,a_4^\pi$ as in \cite{DKMMO}, revealing
a reasonable agreement with the data at   $Q^2\leq 15 $ GeV$^2$.
The most accurate LCSR analysis of the $\gamma^*\gamma \to
\pi^0 $
form factor, including the new twist-6 corrections was carried out
recently in \cite{ABOP}.
It was shown that also the rise of the form factor at high-$Q^2$
observed by the BABAR collaboration \cite{BABARpigam}
(albeit with large errors) can still be accommodated in the LCSR
prediction at the expense
of a moderate ``deformation''
of the shape of $\varphi_\pi(u)$. Most importantly, one
has to increase the coefficient $a_4^\pi(1 \mbox{GeV})$ up to $\sim 0.2$,
leaving $a_2^\pi$ in the ballpark of (\ref{eq:a2a4})
and (optionally) adding small coefficients $a^\pi_{6,8,10}$ to the model.
We illustrate  the numerical influence of
such variation of $\varphi_\pi(u)$ on the $B\to \pi$ LCSR (\ref{eq:LCSR}) in
the next section, and confirm the observation made in \cite{ABOP}
that the heavy-light form factors are only slightly influenced by this
modification of Gegenbauer moments.

The LCSR method can also be used to calculate the $D\to \pi$ form factor.
The recent analysis in \cite{KKMO} where the same intervals
as in \cite{DKMMO} were used
revealed  a very good agreement with
lattice QCD and experiment. We checked that the use of
broader intervals (\ref{eq:a2a4})
does not produce a noticeable effect, simply because the
twist-2 contribution is numerically less important in the $D\to \pi$
LCSR.

\section {Numerical results for the form factors and width}
For the numerical analysis of the LCSR (\ref{eq:LCSR}) and the related
sum rule for $f^0_{B\pi}$ we slightly updated the
input used  in \cite{DKMMO}.
First of all, there is practically no change of the $b$-quark mass.
According to the last update \cite{Chetyrkin}, we adopt
$\bar{m}_b(\bar{m_b})=4.16 \pm 0.03$ GeV, conservatively inflating
the quoted error  by a factor of two.
As explained in detail in \cite{DKMMO}, the $\overline{MS}$-mass
of the $b$-quark is the most suitable mass definition for OPE 
of the correlation function, and the $O(\alpha_s)$ contributions 
to the sum rules are comparably small.
For the $u$- and $d$-quark masses
entering the parameter $\mu_\pi=m_\pi^2/(m_u+m_d)$,
we follow \cite{KKMO} and
use the $s$-quark mass derived from QCD sum rules \cite{ms} and
the ChPT light-quark mass ratios \cite{Leutw}, yielding $[m_u+m_d](2\mbox{GeV})=8.0\pm
1.4$ MeV and, correspondingly
\be
\mu_\pi(2\GeV)=2.43\pm 0.42 \GeV\,,
\label{eq:mupi}
\ee
so that the quark  condensate  density is
$\langle \bar{q}q\rangle(2
\GeV)=-(274^{\,+15}_{\,-17}\,\mbox{MeV})^3$.
This interval is slightly narrower, 
but remains within the broader range used in \cite{DKMMO}. 
As already mentioned in the last section, our intervals for the
parameters $a_2^\pi$ and $a_4^\pi$ given in (\ref{eq:a2a4}) are
broader than the ones used in \cite{DKMMO}.  The rest of
the parameters determining the nonperturbative objects in the sum rules
(DA's and condensate densities), as well as the conventions for the
renormalization and choice of $\alpha_s$ are taken as in \cite{DKMMO}.
As shown there, all nonasymptotic and three-particle 
contributions of the twist-3 DA's as well as the whole 
twist-4 contribution to LCSR are very small, and the uncertainties
in their parameters do not produce visible changes in the
numerical predictions.

The ``internal'' input parameters of our calculation
include the renormalization
scale $\mu$, the Borel parameters $M$ and $\overline{M}$, the
related duality thresholds $s_0^B$ and $\overline{s}_0^B$ in LCSR
and in the 2-point sum rule for $f_B$, respectively. Here we
follow the same strategy as in \cite{DKMMO}, balancing  between
the smallness of the subdominant contributions (twist-4 and
twist-2,3 NLO terms) and a reasonable suppression of the
integrals over the higher states estimated in the quark-hadron
duality approximation. The only minor difference with respect to
the analysis presented in \cite{DKMMO} is that  here we stay on a
more conservative side, allowing for a slightly larger deviation
(up to 3\%) of the calculated $B$-meson mass from its experimental
value. This leads to broader intervals for the Borel parameters
and duality thresholds. More specifically, we adopt the same default
renormalization scale $\mu=$ 3 GeV as in \cite{DKMMO}, allowing
its variation from 2.5 to 4.5 GeV, and use in LCSR the Borel
parameter range  $M^2=(12.0-20.0)\GeV ^2$, with the
threshold parameter gradually shrinking from the interval
$s_0^B=37.5\pm 2.5 \GeV^2$ at $M^2=12.0 \GeV^2$ to
the point $s_0^B=40.0\GeV ^2$ at $M^2=20.0 \GeV^2$.
In the two-point sum rule we vary $\overline{M}^2$ from $ 4.0
\GeV^2~(\overline{s}_0^B=36.5\pm 2.5\GeV^2)$ to $6.0 \GeV^2
~(\overline{s}_0^B=39.0  \GeV^2)$.

\begin{figure}[h]
\centering
\includegraphics[scale=0.60]{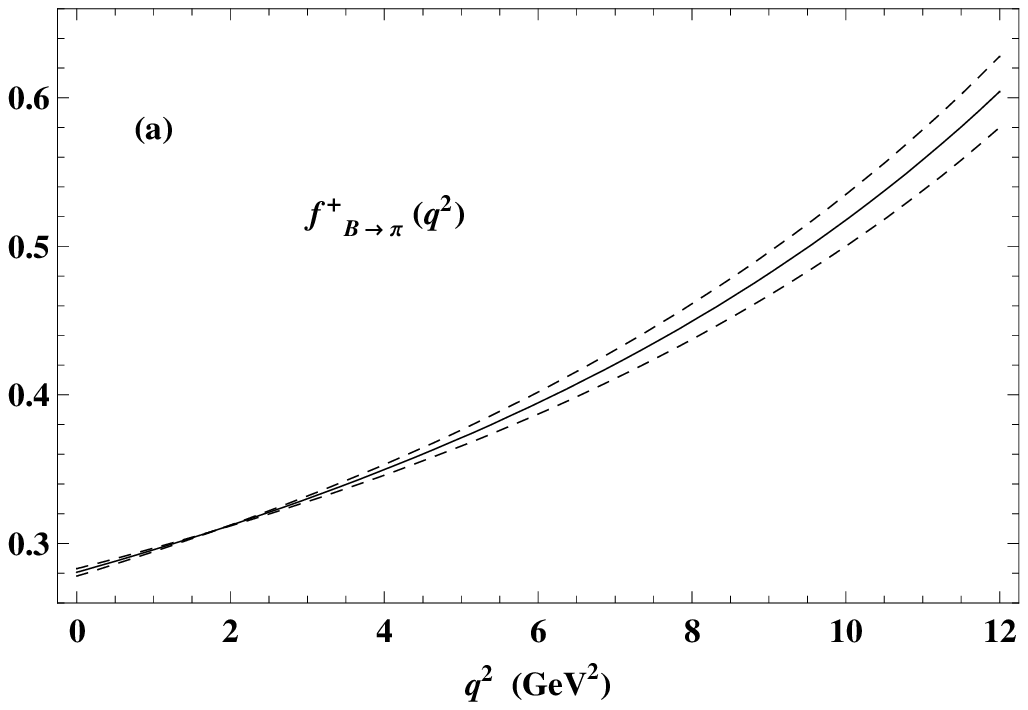}
\includegraphics[scale=0.60]{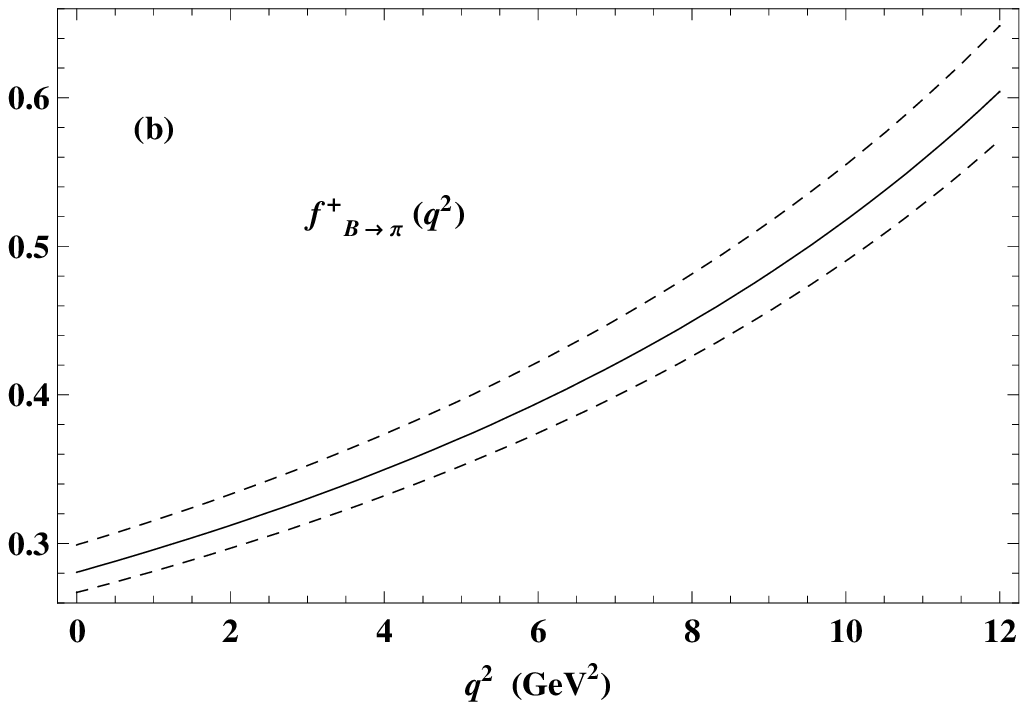}\\
 \includegraphics[scale=0.60]{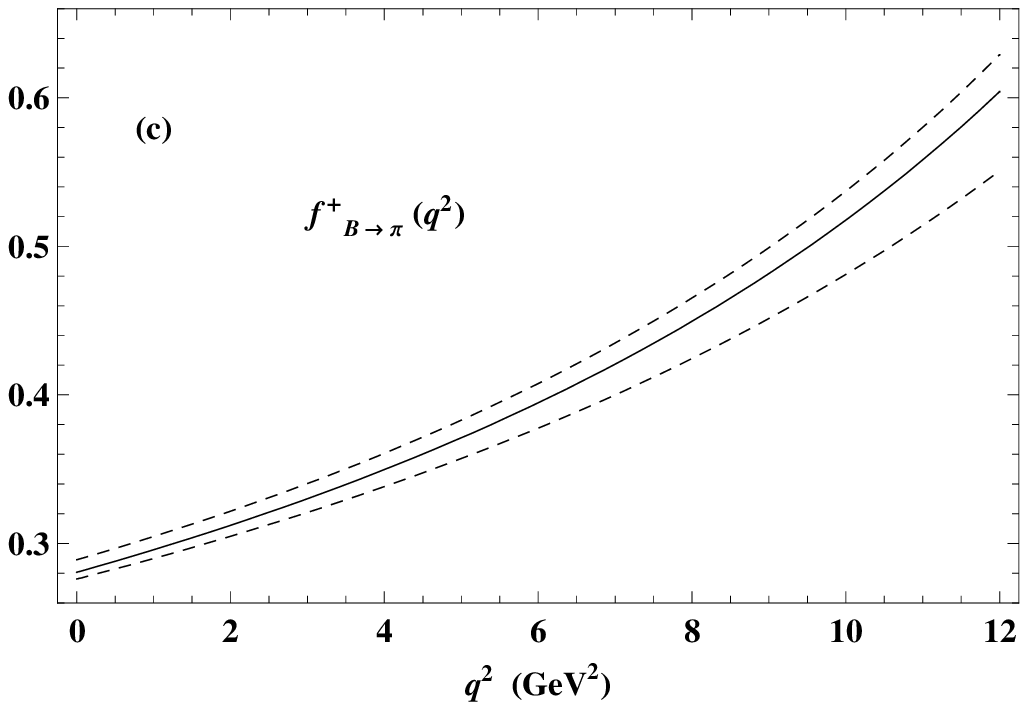}
\includegraphics[scale=0.60]{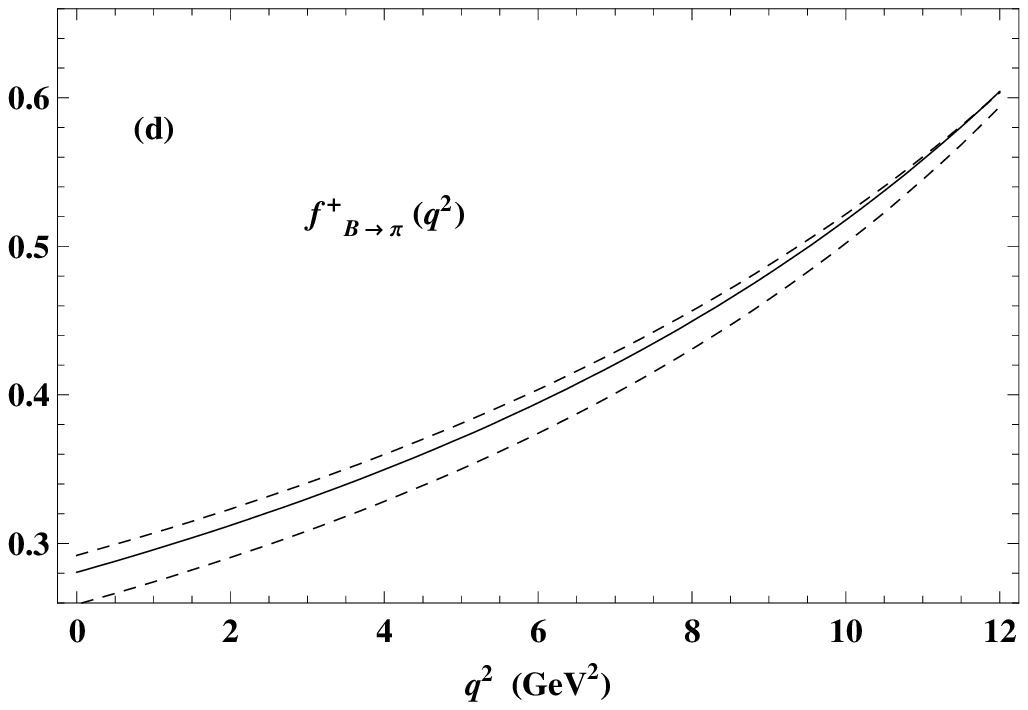}\\
\includegraphics[scale=0.60]{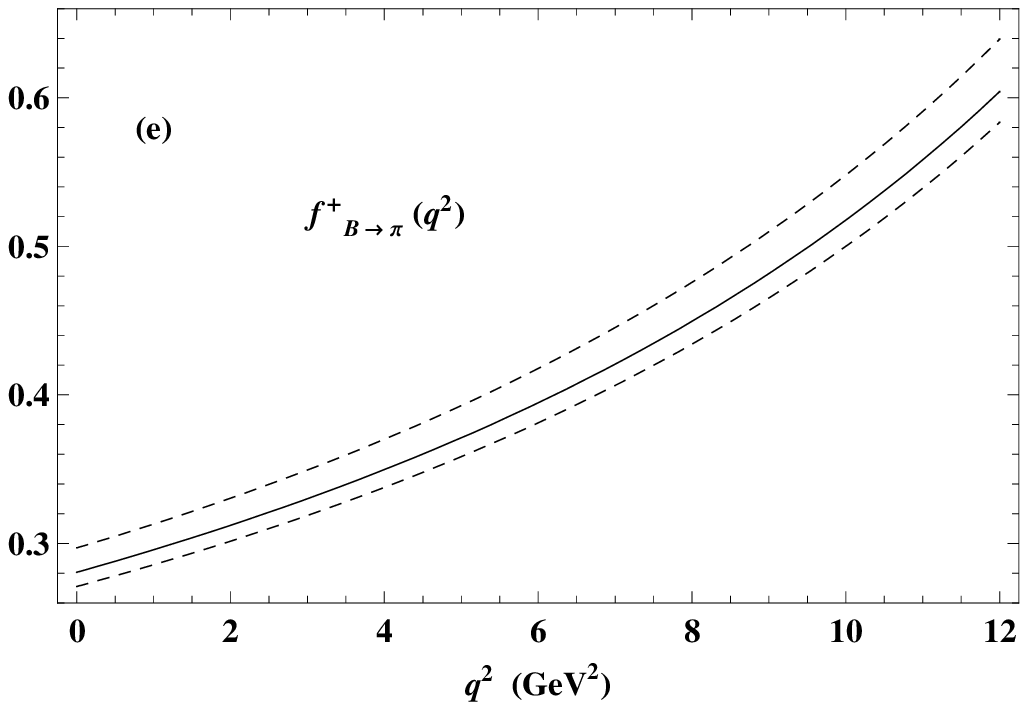}
\includegraphics[scale=0.60]{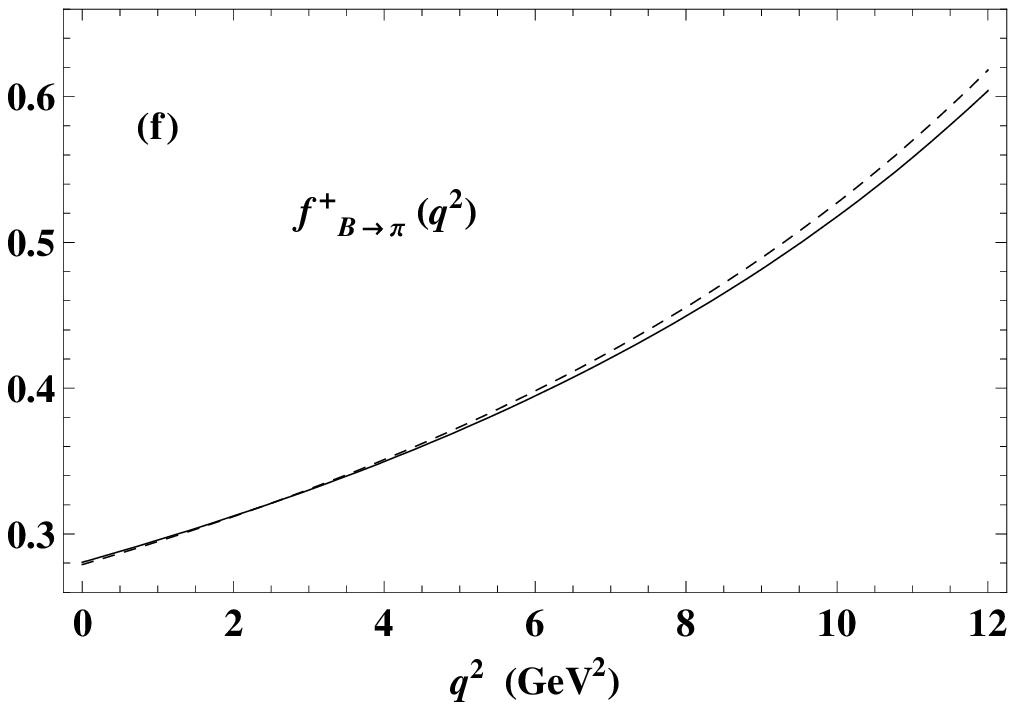}
\vspace{-0.1cm}
\caption{
\it The form factor $f^{+}_{B\pi}(q^2)$
calculated  from LCSR with the central input (solid)
and varying separate input parameters (dashed):
(a) $a_2^\pi,a_4^\pi$, (b) $\mu_\pi$,  (c) $\mu$, (d) $M^2,s_0^B$
and (e) $\overline{M}^2,\overline{s}_0^B$; in (f)
the result for the model III of Gegenbauer moments from \cite{ABOP}
is displayed (dashed).}
\label{fig:fplBpi_uncert}
\end{figure}

The results of our calculation for $f^+_{B\pi}(q^2)$ at $0<\!q^2\!<12.0\GeV^2$
are shown in Fig.~\ref{fig:fplBpi_uncert},
displaying the separate uncertainties
caused by the variation of (a)~$a^\pi_2,a^\pi_4$,
(b)~$\mu_{\pi}$, (c)~$\mu$, (d)~\{$M^2,s_0^B$\} and
(e)~\{$\overline{M},\overline{s}_0^B$\}
within the limits specified above.
In addition, in Fig.~\ref{fig:fplBpi_uncert}f the
default central values of the Gegenbauer moments in (\ref{eq:a2a4})
are replaced by a model with larger $a_4^\pi(1\GeV)=0.22$
(model III in \cite{ABOP}). The small deviation of the form factor
remains within our estimated uncertainty due to the $a^\pi_2,a^\pi_4$
variation.
Since we use a narrow interval of the $b$-quark mass
from \cite{Chetyrkin},
the uncertainties caused by the variation of $\bar{m}_b$ are very small
and not even visible on the plot, hence we do not show them;
the remaining parameters of DA's and condensate densities
generate negligibly small changes of the calculated form factors.
The sensitivity to the renormalization scale is relatively large at
$q^2$ approaching $q^2_{max}$, and also the uncertainties due to
the Borel parameter and duality threshold are now more pronounced
than in \cite{DKMMO} due to an enlargement of the adopted
intervals, whereas the uncertainty due to the twist-3 normalization
$\mu_\pi$ (related to the $u,d$ quark masses and quark condensate)
decreases.
The numerical results for the  form factors \fplq and \fzeroq are consistent with what
was obtained in \cite{DKMMO}; a few percent shift of the
central value of $f^+_{B\pi}(0)$ (presented in Table~1 below) can
be traced to the modification of the input.

Note that the variations of the form factors calculated from LCSR
at different $q^2$ are strongly correlated, in particular,  the
shape of the form factor \fplq is correlated with its value at $q^2=0$.
Quoting separate theoretical errors for each point of accessible
$q^2$-region, including all these correlations makes
the numerical predictions too complex and in fact is not necessary,
since our main goal is the integrated semileptonic width over this region.
Instead, we calculate the deviations
of this width with respect to  the variations of individual input parameters,
so that the correlations are automatically taken into account
after the integration over $q^2$.
Our main result is the integral (\ref{eq:zeta}) calculated
using the LCSR results for $f_{B\pi}^+(q^2)$:
\ba \Delta\zeta(0,12
\GeV^2)&=&4.59 \,\,{}^{+0.16}_{-0.16}\Big|_{a2,a4}
\,\,{}^{+0.03}_{-0.03}\Big|_{m_b}
\,\,{}^{+0.68}_{-0.46}\Big|_{\mu_\pi}
\,\,{}^{+0.31}_{-0.39}\Big|_{\mu}
\,\,{}^{+0.29}_{-0.47}\Big|_{M,s_0}
\,\,{}^{+0.59}_{-0.32}\Big|_{\bar{M},\bar{s}_0}
\mbox{ps}^{-1}\nonumber\\
&=&4.59^{+1.00}_{-0.85}~\mbox{ps}^{-1}\,,
\label{eq:deltaZ}
\ea
where the negligibly small uncertainties related to the rest of the
input are not shown but included in the total error obtained by adding all separate uncertainties in  quadrature
\footnote{~This replaces our
preliminary result $\Delta\zeta (0,12 \GeV^2)=4.00^{+1.01}_{-0.95}$
quoted in \cite{BaBarnew1,Babarnew2} and obtained
with exactly the same input as in \cite{DKMMO}, except no  data
on $B\to \pi \ell \nu_\ell $ were used  and broader
intervals $a_2^\pi(1 \GeV^2)=0.25\pm 0.15$ and $a_4^\pi(1 \GeV^2)=(0.1 \pm
0.1)- a_2^\pi(1 \GeV^2)$ were adopted.
}.
Importantly, (\ref{eq:deltaZ})
has a slightly smaller overall uncertainty than the values
of the form factor \fplq at separate $q^2$, due to the abovementioned
correlations.

Using  (\ref{eq:deltaZ}), we employ the recent BABAR data for the
$B\to \pi \ell \nu_\ell$ width. The  branching fraction integrated
from $q^2=0$ to $q^2=12\GeV^2$ was measured by the BABAR
collaboration using two different techniques, and the results are:
\ba \Delta {\cal B}(0,12\GeV^2)=(0.84\pm 0.03\pm 0.04)\times
10^{-4}~~  \cite{Babarnew2}\,,
\nonumber\\
\Delta {\cal B}(0,12 \GeV^2)=(0.88\pm 0.06)\times 10^{-4} ~~ \cite{BaBarnew1}.
\ea
Taking their weighted average, the
total lifetime $\tau_{B^0}=1.525\pm 0.009$ ps and  substituting
(\ref{eq:deltaZ}) in (\ref{eq:zeta}), we obtain:
\be
|V_{ub}|=\bigg(3.50^{+0.38}_{-0.33}\Big|_{th.}\pm 0.11 \Big|_{exp.}\bigg)
\times 10^{-3}\,,
\label{eq:ourVub}
\ee
where the theoretical error corresponds to the estimated
total uncertainty in (\ref{eq:deltaZ}).

\section{Accessing the large $q^2$  region with
z-parameterization}

\begin{figure}[t]
\begin{center}
\includegraphics[scale=0.90]{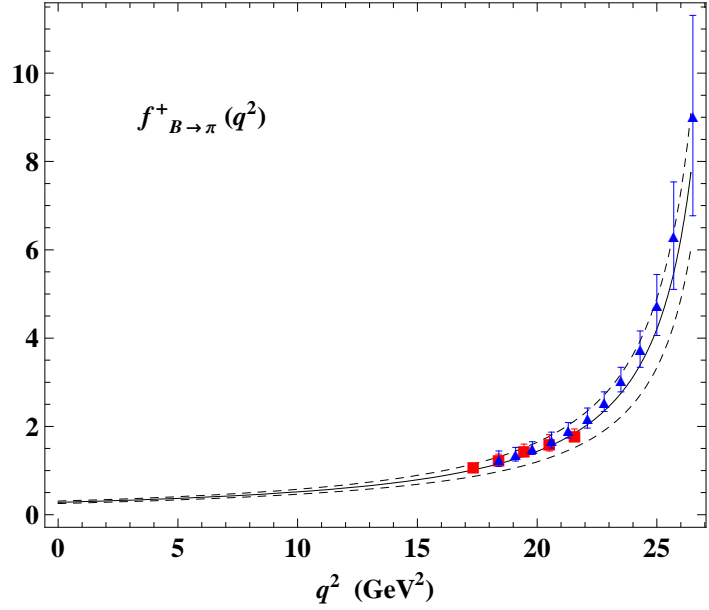}\\
\caption{ \it The vector form factor $f^{+}_{B \pi}(q^2)$
calculated from LCSR and fitted to the BCL parameterization
(solid) with uncertainties (dashed), compared with the HPQCD
\cite{HPQCD} (squares) and FNAL/MILC \cite{FermilabMILC}
(triangles)
 results. } \label{fig-fplusq2}
\end{center}
\end{figure}

To extrapolate the calculated form factor, we use the
$z$-series parameterization
(see e.g., \cite{BCL,BGL})
based on  the analyticity of the form factors and using
the transformation:
\begin{equation}
 z(q^2,t_0)=\frac{\sqrt{(m_B+m_{\pi})^2-q^2}-\sqrt{(m_B+m_{\pi})^2-t_0}}{\sqrt{(m_B+ m_{\pi})^2-q^2}+\sqrt{(m_B+ m_{\pi})^2-t_0}},
 \label{eq:z}
\end{equation}
where
$t_0=(m_B+m_{\pi})^2-2\sqrt{m_Bm_\pi}\sqrt{(m_B+m_{\pi})^2-q^2_{min}}$ is the
auxiliary parameter, chosen to maximally reduce the interval of
$z$ obtained after the mapping (\ref{eq:z}) of the
region $q^2_{min}<q^2<q^2_{max}$, where the LCSR calculation is
valid. More specifically, we adopt  the BCL version \cite{BCL} of
this parameterization, that is, for the vector form factor:
\begin{equation}
 f^+_{B\pi}(q^2) = \frac{1}{1-q^2/m_{B^*}^2}\sum\limits_{k=0}^{N}
\widetilde{b}_k\,[z(q^2,t_0)]^k\,.
\label{eq:BCL}
\end{equation}
As explained in \cite{BCL}, this parameterization
has certain advantages with respect to the
BGL-version \cite{BGL}.
Furthermore, to obey the expected near-threshold behavior, the
relation \be \widetilde{b}_N =
-\frac{(-1)^N}{N}\sum\limits^{N-1}_{k=0}(-1)^k\,k\,
\widetilde{b}_k \label{eq:BCLrel} \ee is implemented, reducing the
number of independent parameters by one.
In addition, we find it more convenient
to keep the form factor at zero momentum transfer $f^+_{B\pi}(0)$
as one of the fit parameters, correspondingly
rescaling the coefficients in the $z$-series expansion.
This leads to the same parameterization
of the vector form factor as the one used in \cite{KKMO}:
\ba
f^+_{B\pi}(q^2) = \frac{f^+_{B\pi}(0)}{1-q^2/m_{B^*}^2}
\Bigg\{1+\sum\limits_{k=1}^{N-1}b_k\,\Bigg(z(q^2,t_0)^k-
z(0,t_0)^k
\nonumber\\
-(-1)^{N-k}\frac{k}{N}\bigg[z(q^2,t_0)^N-
z(0,t_0)^N\bigg]\Bigg)\Bigg\}
\label{eq:BCLfpl}\,.
\ea

\begin{figure}[t]
\begin{center}
\includegraphics[scale=0.90]{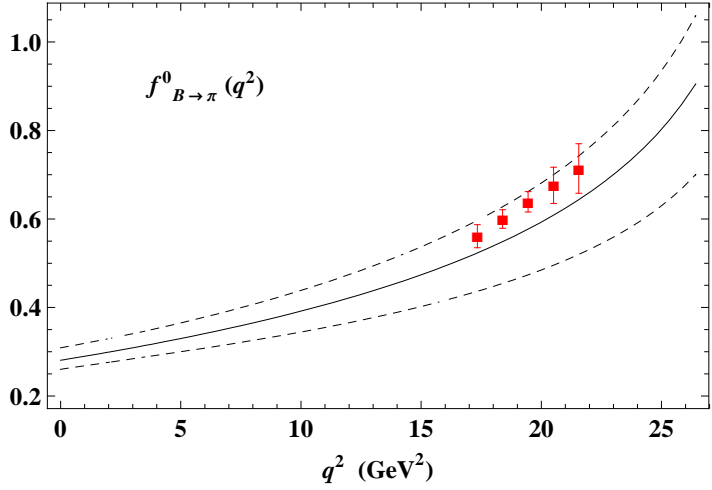}\\
\caption{ \it The scalar form factor $f^{0}_{B\pi}(q^2)$
calculated from LCSR and fitted to the BCL
parameterization. The notations are the same as in Fig.~ \ref{fig-fplusq2}.}
\label{fig-f0q2}
\end{center}
\end{figure}

\begin{figure}[t]
\begin{center}
\includegraphics[scale=1.0]{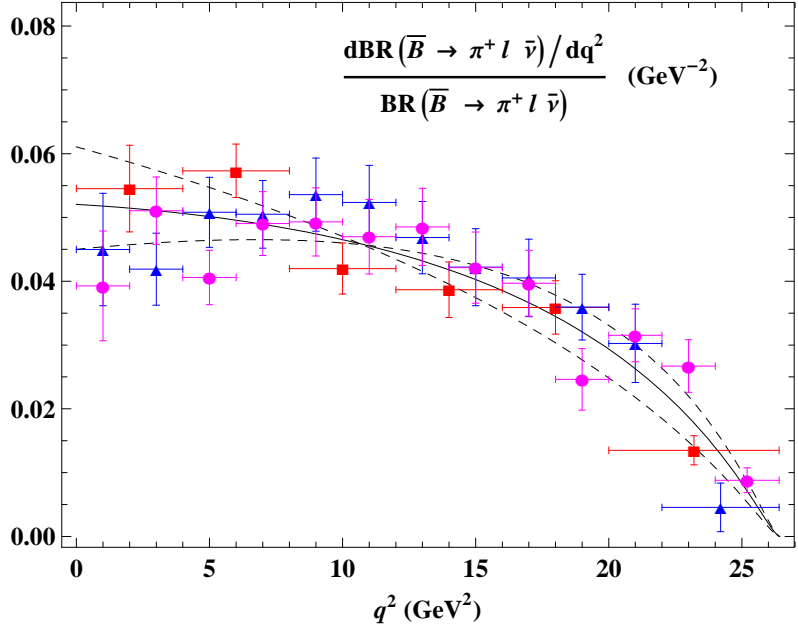}\\
\caption{ \it (colour online) The normalized $q^2$-distribution in $B \to \pi l \nu$ obtained from LCSR and
extrapolated with the $z$-series parameterization (central input- solid,
uncertainties -dashed). The experimental data points are
from BABAR: (red) squares \cite{BaBarnew1}, (blue) triangles \cite{Babarnew2}
and Belle \cite{Belle}: (magenta) full circles.}
\label{fig-Elept}
\end{center}
\end{figure}

The scalar form factor $f^{0}_{B\pi}(q^2)$ is
parameterized in a similar way, except that there is no pole factor
for the obvious reason: the lowest $B$-resonance in the
$J^P=0^+$ channel
is located above the $B\pi$ threshold. Thus, we use:
\ba
f^0_{B\pi}(q^2) = f^0_{B\pi}(0)
\Bigg\{ 1+\sum\limits_{k=1}^{N} b_k^0\,
\Bigg(z(q^2,t_0)^k- z(0,t_0)^k\Bigg)\Bigg\}
\label{eq:BCLf0}\,,
\ea
where by default $f^0_{B\pi}(0) = f^+_{B\pi}(0)$.

We fitted the numerical LCSR prediction for the form factor
\fplq to (\ref{eq:BCLfpl})
with $N=2$
and and $f^0_{B\pi}(q^2)$ to (\ref{eq:BCLf0}) with $N=1$, respectively.
To increase the ``lever arm'' we also employed the LCSR
predictions at negative $q^2$, up to $q^2_{min}=-6.0 \GeV ^2$.
After the mapping (\ref{eq:z}), $q^2_{min}\to z=0.30$ and
$q^2_{max}=12 \GeV^2\to z=0.13$, so that the values of $z$ are sufficiently
small  to justify truncating the expansion (\ref{eq:BCL}).
The number $N$ of terms in this expansion can be made larger,
with no essential change in the fitted form factor but with increasing
individual uncertainties for the coefficients $b_k$
and with large correlations between them.  We checked that
the upper bounds on the expansion coefficients
$b_k$ following from the OPE of the two-point correlation function
of the vector $\bar{b}\gamma_\mu u $ currents
(see \cite{BCL} for detailed expressions) are far from being saturated
for low $N$. We also used the analogous bounds
for the scalar form factor  obtained recently in \cite{BFW}.
Altogether, the OPE bounds  play a role starting from $N=5$.

The fitted values of $f^{+}_{B\pi}(0)=f^0_{B\pi}(0)$  and of the
slope parameters $b_1,b_1^0$ are presented in Table~\ref{tab:fitres},
together with  the numerically important uncertainties,
the latter revealing significant correlations.
With these results we extrapolate
the form factors at $q^2>q^2_{max}$ and compare with the
lattice QCD results.
This is shown in Fig.~\ref{fig-fplusq2} for \fplq and in
Fig.~\ref{fig-f0q2} for \fzeroq \!.

\begin{table}[h]
\begin{center}
\begin{tabular}{|c|c|c|c|c|c|c|}
  \hline
&&&&&& \\[-3.5mm]
 Parameter & centr. value & \{$a_2, a_4$\}  & $\mu_{\pi}$ & $\mu$ &
$\{M^2, s_0\}$ &
$\{\overline{M}^2, \overline{s}_0\}$ \\
  \hline
    &&&&&&\\[-1mm]
  $f^+_{B\pi}(0)$ &$0.281$ & $^{+0.002}_{-0.003}$&$^{+0.018}_{-0.014}$&
$^{-0.005}_{+0.008}$&
$^{+0.010}_{-0.022}$ & $^{-0.010}_{+0.016}$ \\
  &&&&&& \\[-2mm]
\hline
&&&&&& \\[-3mm]
  $b_1$ & $-1.62$ & $^{+0.43}_{-0.44}$ & $^{-0.06}_{+0.05}$ &
$^{+0.53}_{-0.07}$ & $^{+0.30}_{-0.49}$ &-\\[2mm]
\hline
&&&&&& \\[-2mm]
   $b^{0}_1$ &$-3.98$  & $^{+0.56}_{-0.57}$ & $^{-0.28}_{+0.23}$ &
$^{+0.96}_{-0.08}$  & $^{+0.28}_{-0.42}$ &-\\[2mm]
    \hline
\end{tabular}
\end{center}
\caption{ \it Fitted parameters for $z$-series parameterization
of the  form factors $f^{+,0}_{B
\pi}(q^2)$ and their uncertainties due to the variations of
the input parameters.}
\label{tab:fitres}
\end{table}

The dashed curves in these figures are obtained by
adding separate variations of the form factors
in quadrature at each $q^2$, so that the variations
of the solid curves corresponding to the central input
are bounded within the area between the upper and lower dashed curves.
As expected, the uncertainties of the form factors
extrapolated to larger $q^2$, exceed the ones calculated at smaller $q^2$.
This circumstance,
however, does not play a significant role for the integrated widths,
since the integration over the phase space
suppresses the semileptonic width in the large $q^2$-region.

Our predictions  for \fpl are, within errors,
in a reasonable agreement with the lattice QCD results obtained
by HPQCD \cite{HPQCD} and Fermilab/MILC \cite{FermilabMILC} collaborations.
We also observe an agreement with the
normalization and shape of the form factors obtained by the
QCDSF collaboration \cite{QCDSF}, in particular,
they predict $f^+_{B\pi}(0)=0.27\pm 0.07\pm 0.05 $.

Furthermore,  we estimate the  total width of $B\to \pi \ell \nu_\ell$
in units of $1/|V_{ub}|^2$ and the integral (\ref{eq:zeta}) for the large $q^2$-region:
\ba
\frac{1}{|V_{ub}|^2}\Gamma(B\to \pi \ell \nu_\ell)
=\Delta \zeta (0,26.4 \GeV^2) =7.71
^{+1.71}_{-1.61}~\mbox{ps}^{-1}\,,\nonumber\\
\Delta \zeta (16 \GeV^2, 26.4
\GeV^2)=1.88^{+0.53}_{-0.59}~\mbox{ps}^{-1}\,. \label{eq:totwidth}
\ea

Our prediction for the latter integral has to be compared
with the lattice QCD results presented below, in Table~\ref{tab:2}.

In Fig.~\ref{fig-Elept} we plot the predicted $q^2$-shape in $B\to
\pi \ell \nu_\ell$  obtained by calculating  the normalized
differential width $(1/\Gamma)d\Gamma/dq^2$ from LCSR at $0\leq
q^2 \leq q^2_{max}$ and from the $z$-series parameterization at
$q^2_{max}<q^2\leq (m_B-m_\pi)^2$. The estimated uncertainties are
naturally smaller than in Fig~\ref{fig-fplusq2}, because the
variations of the form factor normalization cancel in the ratio of
the differential and total widths. Our result is compared with the
measured $q^2$-distributions. We use the partial $\Delta{\cal B}$
spectrum obtained by the BABAR collaboration in 6 bins  and 12
bins from  the two independent analyses \cite{BaBarnew1} and
\cite{Babarnew2}, respectively. For the  normalization we employ
the corresponding central values of the measured total branching
fractions:  ${\cal B}(B^0\to \pi^-\ell^+\nu_\ell)= (1.41\pm
0.05\pm 0.07)\times 10^{-4}$\cite{BaBarnew1} and ${\cal B}(B^0\to
\pi^-\ell^+\nu_\ell)= (1.42\pm 0.05\pm 0.07)\times 10^{-4}$
\cite{Babarnew2}. The analogous 13-bin distribution measured by
the Belle collaboration \cite{Belle} and normalized by their total
branching fraction ${\cal B}(B^0\to \pi^-\ell^+\nu_\ell)= (1.49\pm
0.04\pm 0.07)\times 10^{-4}$ is also shown. Fig.~\ref{fig-Elept}
reveals a general agreement of our prediction for the $q^2$-shape
and experimental results, however, only within still large
uncertainties of both experiment and theory.  In particular, the
shape of the form factor fitted from the BABAR data in
\cite{BaBarnew1}  to the same BCL parameterization with two
parameters yields: $(\tilde{b}_1/\tilde{b_0})_{BABAR}=-0.67\pm
0.18$, whereas  our result for the same ratio is
$(\tilde{b}_1/\tilde{b_0})_{LCSR}=-1.10^{+0.40}_{-0.27}$~.

\begin{figure}[h]
\begin{center}
\includegraphics[width=10cm]{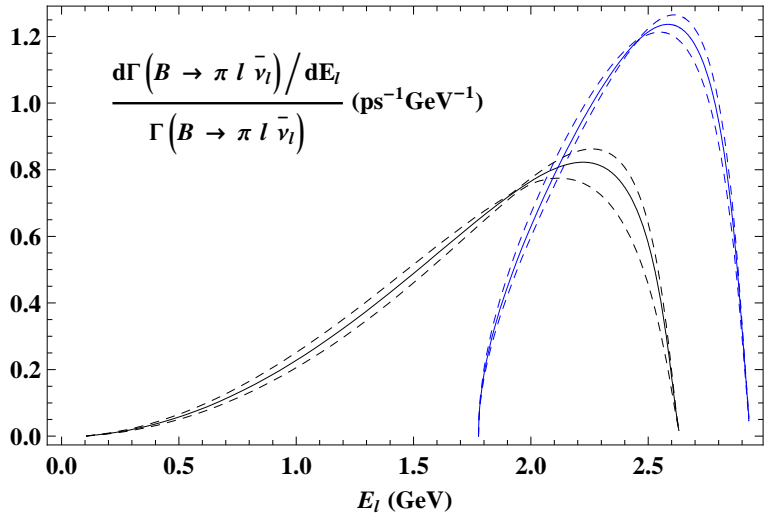}\\
\caption{ \it Lepton energy spectra for $B \to \pi \ell \nu_{\ell}$ at
$m_\ell\leq E_\ell\lesssim(m_B^2+m_\ell^2)/(2m_B)$ for $\ell=\mu,\tau$.
Solid (dashed) lines correspond to the
form factors calculated at the central input (indicate the uncertainties).}
\label{fig-lepenerg}
\end{center}
\end{figure}

A ``byproduct'' observable that was not yet measured
in $B\to \pi \ell \nu_\ell$, is the distribution of the
lepton energy in the $B$-meson rest frame
shown in Fig.~\ref{fig-lepenerg} in the normalized form.
In the electron or muon semileptonic $B$-decay, at a given lepton
energy, the distribution $d\Gamma(B\to \pi \ell\nu_\ell)/dE_l$
contains the integral of $|$\fplq$|^2$ over the region
$0<q^2\lesssim 2m_BE_l$ (the expression for this distribution, also at
$m_\ell \neq 0$, can be found, e.g., in \cite{KR}).
Hence the correlations
between the normalization and shape of the form factor somewhat reduce the
uncertainties in this distribution. It also
has a more pronounced slope than the $q^2$ distribution.

\section{$B\to \tau\nu_\tau$ and  $B\to \pi \tau \nu_{\tau}$}

Currently, the  leptonic width  $B\to \tau \nu_\tau$
measured  by both BABAR
 and Belle collaborations
(see Table 2) is larger than the SM prediction:
\be
{\cal B}(B^-\to \tau\bar{\nu}_\tau)=
\frac{G_F^2}{8\pi}|V_{ub}|^2 m_\tau^2 m_{B}
\left(1-\frac{m_\tau^2}{m_B^2}
\right)^2f_B^2\tau_{B^-}\,,
\label{eq:Btaunuwidth}
\ee
if one employs $f_B$ predicted  from lattice QCD
or QCD sum rules, together with
$|V_{ub}|$ extracted from $B\to \pi\ell\nu_{\ell}$.
The recent discussions on this situation
are mostly concentrated on the value of $|V_{ub}|$. Indeed,
the tension decreases, if one uses
in ({\ref{eq:Btaunuwidth}) the somewhat larger value of
$|V_{ub}|$ extracted from the inclusive $b\to u$ decays.
On the other hand,
the CKM fits \cite{CKMfitter,UTfit} yield a smaller
$|V_{ub}|$, consistent with the determinations from $B\to \pi\ell\nu_{\ell}$.

Let us emphasize that, independent of the actual $|V_{ub}|$  value, there
exists a tension between the ratio of semileptonic and leptonic $B$ widths
and the QCD predictions for the two relevant hadronic matrix elements \fplq and $f_B$.
To demonstrate  that,  we define the following observable:
\be
R_{s/l}(q_1^2,q_2^2)\equiv
\frac{\Delta{\cal B}_{B\to \pi\ell \nu_\ell}(q_1^2,q_2^2)}
{{\cal B}(B\to \tau \nu_\tau)}\left(\frac{\tau_{B^-}}{\tau_{B^0}}\right)
=\frac{\Delta\zeta(q_1^2,q_2^2)}{(G_F^2/8\pi)
m_\tau^2 m_{B}(1-m_\tau^2/m_B^2)^2f_B^2}
\,,
\label{eq:ratio}
\ee
where the partial branching fraction $\Delta{\cal B}$ and the integral
$\Delta\zeta$ defined as in (\ref{eq:zeta}),
are taken over the same region $q_1^2\leq q^2\leq q_2^2$ of the momentum transfer.

The above equation  for the ratio $R_{s/l}$ follows solely from the
$V-A$ structure of the weak currents  in SM
and $V_{ub}$ cancels out in the ratio. The form factor \fpl and decay constant
$f_B$ entering r.h.s. are obtained by one and the same QCD method: lattice QCD
or the combination of LCSR and QCD sum rule.
In Tables~\ref{tab:2} and \ref{tab:3} we collect the inputs
for this equation, obtained from different measurements and
QCD calculations. The disagreement between  the calculated and measured
ratio $R_{s/l}$ goes beyond the theoretical
and experimental errors, especially in the case of
the lattice calculations which have smaller uncertainties.
{\small
\begin{table}[h]
\begin{center}
\begin{tabular}{|c|c|c|c|c|}
\hline
&&&\\[-3mm]
Exp. &$\Delta{\cal B}(10^{-4})$ [Ref.] &${\cal B}(B\to \tau\nu_\tau)(10^{-4})$~~[Ref.]& $R_{s/l}$ \\
&&&\\
\hline
&&&\\[-3mm]
BABAR&$0.32\pm 0.03$  \cite{BaBarnew1}&$1.76\pm 0.49$
\cite{BaBarfB,Barlow}&$0.20^{+0.08}_{-0.05}$\\
&$0.33\pm 0.03\pm 0.03$ \cite{Babarnew2}&&\\
&&&\\
Belle&$0.398\pm 0.03$ \cite{Belle}&  $1.54^{+0.38}_{-0.37}\mbox{}^{+0.29}_{-0.31}$
\cite{BellefB}&$0.28^{+0.13}_{-0.07}$\\[2mm]
\hline
\hline
&&&\\[-1mm]
QCD & $\Delta\zeta$(ps$^{-1}$)~~[Ref.]&$f_B$(MeV) [Ref.] &  $R_{s/l}$ \\
&&&\\
\hline
&&&\\[-4mm]
HPQCD &$2.02\pm 0.55$ \cite{HPQCD}&$190\pm 13$ \cite{HPQCDfB}&$0.52\pm 0.16$\\
&&&\\
FNAL/MILC &$2.21^{+0.47}_{-0.42}$ \cite{FermilabMILC}&$212\pm 9$ \cite{FMILCfB}&
$0.46\pm 0.10$\\[2mm]
\hline
\end{tabular}
\end{center}
\caption{The ratio $R_{s/l}$ for the region $16 \GeV^2<q^2<26.4\GeV^2$,
 measured and calculated from (\ref{eq:ratio}) using
the lattice QCD results.  The weighted
average over the two BABAR measurements is taken and all errors are added
in quadrature.}
\label{tab:2}
\end{table}
\begin{table}[h]
\begin{center}
\begin{tabular}{|c|c|c|c|c|}
\hline
&&&\\[-2mm]
Exp. &$\Delta{\cal B}$$(10^{-4})$~~[Ref.] &${\cal B}(B\to \tau\nu_\tau)(10^{-4})$ [Ref.]& $R_{s/l}$\\
&&&\\
\hline
&&&\\[-3mm]
BABAR &$0.88\pm 0.06$ \cite{BaBarnew1}&$1.76\pm 0.49$ \cite{BaBarfB,Barlow}&$0.52^{+0.20}_{-0.12}$\\
&$0.84\pm 0.03\pm 0.04$ \cite{Babarnew2}&&\\[2mm]
\hline\hline
&&&\\[-3mm]
QCD & $\Delta\zeta $~~[Ref.]&$f_B$(MeV) [Ref.] & $R_{s/l}$\\
&&& \\
\hline
&&&\\[-3mm]
LCSR/QCDSR &$4.59^{+1.00}_{-0.85}$ [this work] &$210\pm 19$
\cite{JL}&
$0.97^{+0.28}_{-0.24}$\\[2mm]
\hline

\end{tabular}
\end{center}
\caption{The same as in Table~\ref{tab:2} for the region $0\leq q^2 \leq 12.0\GeV^2$
where the QCD sum rule results are used.}
\label{tab:3}
\end{table}
}

Decreasing further the theoretical and experimental errors
in (\ref{eq:ratio}),
especially in  the $B\to \tau\nu_\tau$  width, becomes  therefore a very
important task. Possible effects beyond the SM in $B\to \tau\nu_\tau$
are already being discussed  in the literature, and, in particular,
$B\to D\tau\nu_\tau$ is proposed as a channel
which has common  new physics contributions with the leptonic
$B$ decay (see e.g., \cite{NTW} and references therein).

Here we would like to attract attention
to another semileptonic channel: $B\to \pi \tau\nu_\tau$,
although it is experimentally very demanding.
Earlier this channel was discussed e.g., in \cite{KR,DKS}.
Note that this channel has the same
combination of quark and lepton flavours as $B\to \tau \nu_\tau$.
In the SM, the $B\to \pi \tau\nu_\tau$ decay differs
only kinematically from the semileptonic modes with the muon or electron.
A convenient, $V_{ub}$-independent observable  \cite{Khodjamirian:2009zz}
is the ratio
\ba
\frac{d\Gamma(B\to \pi \tau\nu_\tau)/dq^2}{ d\Gamma(B\to \pi
\ell\nu_\ell)/dq^2} =\frac{(q^2-m_\tau^2)^2}{(q^2)^2}
\Big(1+\frac{m_\tau^2}{2q^2}\Big)
\nonumber\\
\!\times
\!\Bigg\{1+\frac{3m_\tau^2(m_B^2-m_\pi^2)^2}{4(m_\tau^2+2q^2)m_B^2p_\pi^2}\frac{|f^0_{B\pi}
(q^2)|^2}{|f^+_{B\pi}(q^2)|^2}\!\Bigg\}\,, \label{eq:Rtau} \ea
where $\ell=e$ or $\mu$ and $m_\ell$ is neglected.
It is determined by the ratio of the scalar and vector
$B\to \pi$ form factors, hence, it has a somewhat smaller
uncertainty than the individual form factors. The ratio
(\ref{eq:Rtau}) is plotted in Fig.~\ref{fig:tauel}
in the kinematically allowed region $m_\tau^2<q^2<(m_B-m_\pi)^2$, using our
predictions for the form factors. It grows at large $q^2$,
due to the kinematical suppression of the vector channel
contribution at small $p_\pi$. The strong correlations between
the vector and scalar form factors calculated from LCSR with
one and the same input  result in a small uncertainty in their ratio.
The tau-lepton energy spectrum
for  $B\to \pi \tau \nu$ shown in Fig.~\ref{fig-lepenerg}
is another observable which depends on both vector and scalar
form factors.

To investigate the influence of new physics on both leptonic
and semileptonic $B$ decays with a $\tau$-lepton,
we include in the effective Hamiltonian of the
$b\to u \tau\nu_\tau$ transition
an intermediate charged Higgs-boson contribution
adopted in the same generic form as in \cite{NTW} :
\ba
H_{eff}=\frac{G_F}{\sqrt{2}}V_{ub}\bigg\{
\bar{u}\gamma_\mu(1-\gamma_5)b~
\bar{\tau}\gamma^\mu(1-\gamma_5)\nu_\tau
\nonumber\\
-\frac{\bar{m}_b m_\tau}{m_B^2}
\bar{u}\bigg( g_S+g_P\gamma_5\bigg) b \bar{\tau}(1-\gamma_5)\nu_\tau \bigg\}
+h.c.\,,
\label{eq:Heff}
\ea
The admixture of
new physics in $B\to \tau \nu_\tau$ and $B \to \pi\tau \nu_\tau$
is then determined, respectively, by the pseudoscalar and scalar
parts of the new interaction.
In particular, the leptonic width (\ref{eq:Btaunuwidth}) gets multiplied by $(1-g_P)^2$;
and therefore, with this choice, the $B\to \tau \nu_\tau$  width vanishes at $g_P=1$.
Accordingly, the r.h.s. of (\ref{eq:ratio}) acquires a factor $1/(1-g_P)^2$.
Also the ratio (\ref{eq:Rtau}) is modified
by multiplying the scalar form factor with an additional factor:
\be
f^0_{B\pi}(q^2) \to \Big(1-\frac{g_S~q^2}{m_B^2}\Big)f^0_{B\pi}(q^2),
\label{eq:f0eff}
\ee

The addition of the new interaction can  fill  the gap
between the calculated and experimentally measured ratio $R_{s/l}$
in (\ref{eq:ratio}) if  one allows for $g_P\neq 0$.
Taking, e.g., the sum rule prediction for
this ratio from Table 3, and adding the new physics contribution,
$R_{s/l}\to R_{s/l}/(1-g_P)^2$, we equate it to the experimental value and
find that $g_P\neq 0 $ is allowed within one of the following
two intervals: $g_P=-(0.4\pm 0.2|_{th}\pm 0.2|_{exp})$ or
$g_P=2.4\pm 0.2|_{th}\pm 0.2|_{exp}$.
Assuming the parameter $g_S$  in the same
ballpark as $g_P$ (in fact, $g_{S}=g_P$ in MSSM, however, only with
positive values), we display
in Fig.~\ref{fig:rtau} the modified ratio (\ref{eq:Rtau}),
adding the new physics contribution with $g_S=-0.4$
and $g_S=2.4$. In fact, the second option may already be
excluded by the $B\to D \tau \nu_\tau$ analysis (see e.g.,
\cite{2Higgs}). In addition, in Fig.~\ref{fig:tauel}, the ratio of total semileptonic widths is plotted as a function
of $g_S$. Note that the deviation due to new physics can be larger than
the uncertainty due to the hadronic form factors.

\begin{figure}
\begin{center}
\includegraphics[width=10cm]{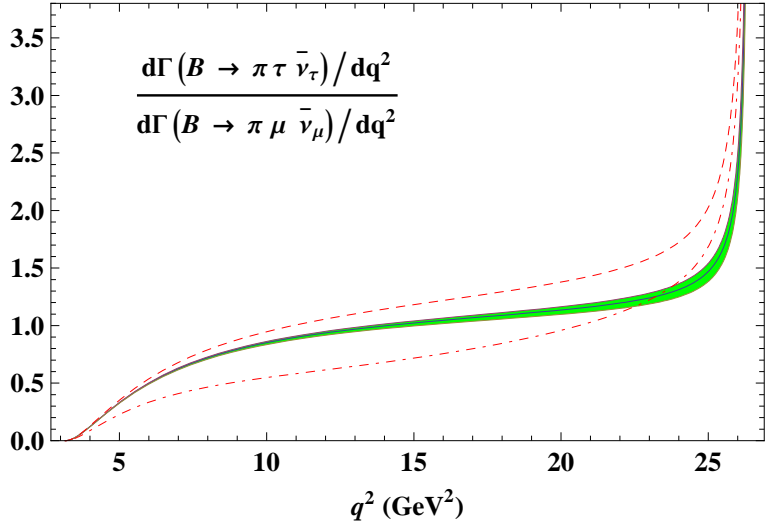}\\
\caption{ \it (colour online) Ratio of differential decay widths, defined in
(\ref{eq:Rtau}) (solid), with shaded (green) area indicating the uncertainties.
Also shown is the effect of adding a charged Higgs-boson contribution
with $g_S=-0.4$ (dashed, red) and $g_S=2.4$(dash-dotted,red)}
\label{fig:rtau}
\end{center}
\end{figure}
\begin{figure}
\begin{center}
\includegraphics[width=10cm]{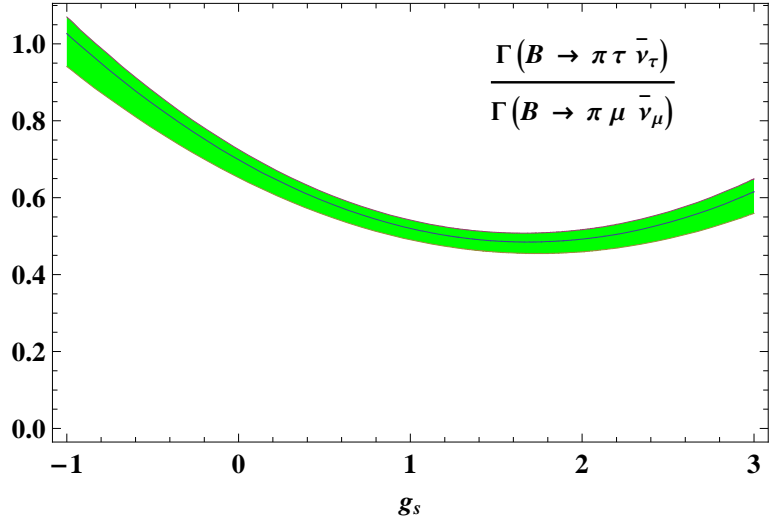}\\
\caption{ \it Ratio of total semileptonic decay widths into $\tau$ and $\mu$(or $e$)
(solid), with shaded (green) area indicating the uncertainties shown as a function
of the parameter $g_S$ determining the charged Higgs-boson coupling in $B\to \pi \tau\nu_\tau$}
\label{fig:tauel}
\end{center}
\end{figure}
\section{Discussion}

The precise determination of the CKM matrix element $V_{ub}$ is mandatory
for stringent tests of the quark-flavour content of SM. Due to recent work in lattice QCD and in QCD sum rules, combined with constraints from
analyticity and unitarity, the values for $|V_{ub}|$ extracted from $B \to \pi \ell \bar{\nu}$ are becoming quite
precise. The main result of this paper is the prediction for the partial width of this decay in the $q^2$-region
$0 \leq q^2 \leq 12 \GeV^2$ in terms of $1/|V_{ub}|^2$, which is expressed as a weighted
integral $\Delta\zeta(0,12 \GeV^2)$ over the squared form factor \fplq.
This integral has smaller uncertainties than the values of the form
factor at separate $q^2$ and allows the $|V_{ub}|$ extraction with
an accuracy approaching 10\%, somewhat better than in the
previous LCSR analysis \cite{DKMMO} where only the value of
the form factor at $q^2=0$ was used.
We extract a value
for $|V_{ub}|$ using recent BABAR data \cite{BaBarnew1,Babarnew2}.
We hope that also the Belle collaboration will
in future provide  the integrated width in this region.

We also employed the $z$-series parameterization with BCL-ansatz
and extrapolated the LCSR form factors to the whole kinematic
region. The $B\to \pi$ form factors \fplq and \fzeroq obtained
from LCSR are, within comparable uncertainties, in agreement with
the recent lattice QCD results and with the measured $q^2$-shape
of the $B\to \pi \ell \nu_\ell$ width. A combined fit of the form
factors calculated at small $q^2$ from LCSR and at large $q^2$
from lattice QCD to a common $z$-series parameterization is
possible (see e.g. the previous analysis in \cite{BCL}) but is
beyond the scope of this paper.

A further improvement would need more precise measurements of the
shape to control the input parameters used in LCSR. On the
theoretical side, the renormalization scale dependence can be
reduced by including NNLO corrections to the hard scattering
amplitudes, and also by separating the renormalization and the
factorization scales. Improving the duality  approximation is more
difficult and demands a knowledge of radially excited states in
$B$ channel.  Another perspective is a simultaneous global fit of
three different LCSR's to the data on $B\to \pi \ell \nu_\ell$ and
on the pion electromagnetic and transition form factors, with
scanning over the allowed region of input.

The value for $|V_{ub}|$ obtained from $B\to \pi \ell \nu_{\ell}$ is somewhat lower
than what is extracted from inclusive $B$ decays, as well
as the one extracted from $B \to \tau \bar{\nu}$, but the significance
is still too small to
be conclusive. On the other hand, our result is completely compatible with
the results from  the CKM fits \cite{CKMfitter,UTfit}.

The impact of the recent measurements of
$B \to \tau \bar{\nu}$ on $V_{ub}$ has not
improved the situation concerning our knowledge of this quantity.
However, the ratio of leptonic
and semileptonic
widths is independent of $V_{ub}$ and may either be
regarded as a test for a possible non-standard
contribution (like, e,g. a charged Higgs-boson exchange) or as a
test of our understanding of QCD. In turn,
a lack of understanding of $f_B$ and the form factor
\fpl will severely limit the sensitivity to a ``new
physics'' contribution.  It is interesting to note that both QCD sum rules as well as lattice calculations
tend to yield larger values of a suitably defined ratio of semileptonic versus leptonic
widths.
This means that we have either a problem in our understanding of QCD matrix
elements or that there
is really a substantial new-physics contribution in  $B \to \tau \bar{\nu}$,
which also makes the channel $B\to \pi \tau \nu_\tau$ very interesting.
A significant test of these
statements has to await more data, in particular on $B \to \tau \bar{\nu}$.

\section* {Acknowledgments}

We are grateful to Martin Jung and Christoph Klein for useful comments
and acknowledge valuable discussions concerning the BABAR collaboration
data with Jochen~Dingfelder and Paul~Taras.
This work is supported by the German research foundation
DFG under the contract  No. KH205/1-2 and by the German Ministry
of Research (BMBF), contract 05H09PSF.

\end{document}